\documentclass[aps,prd,twocolumn,showpacs,superscriptaddress]{revtex4}

\usepackage{color}
\usepackage{graphicx}
\usepackage{epsfig}
\usepackage{psfrag}
\usepackage{bbm}
\usepackage{rotating}
\usepackage{multirow}
\usepackage{amsmath}

\begin{document}

\title{Double optical spring enhancement for gravitational wave detectors}

\author{Henning Rehbein}
\author{Helge M\"uller-Ebhardt}
\affiliation{Max-Planck-Institut f\"ur Gravitationsphysik
             (Albert-Einstein-Institut), \\ Institut f\"ur Gravitationsphysik,
              Leibniz Universit\"at Hannover, Callinstr. 38, 30167 Hannover,
              Germany}
\author{Kentaro Somiya}
\affiliation{California Institute of Techology, Theoretical Astrophysics 130-33, Pasadena, California 91125, USA}
\affiliation{Max-Planck-Institut f\"ur Gravitationsphysik
(Albert-Einstein-Institut),
             Am M\"uhlenberg 1, 14476 Potsdam, Germany}
\author{Stefan L. Danilishin}
\affiliation{Department of Physics, Moscow State University,
Moscow 119992, Russia}
\affiliation{Max-Planck-Institut f\"ur Gravitationsphysik
             (Albert-Einstein-Institut), \\ Institut f\"ur Gravitationsphysik,
              Leibniz Universit\"at Hannover, Callinstr. 38, 30167 Hannover,
              Germany}
\author{Roman Schnabel}
\author{Karsten Danzmann}
\affiliation{Max-Planck-Institut f\"ur Gravitationsphysik
             (Albert-Einstein-Institut), \\ Institut f\"ur Gravitationsphysik,
              Leibniz Universit\"at Hannover, Callinstr. 38, 30167 Hannover,
              Germany}
\author{Yanbei Chen}
\affiliation{California Institute of Techology, Theoretical Astrophysics 130-33, Pasadena, California 91125, USA}
\affiliation{Max-Planck-Institut f\"ur Gravitationsphysik
(Albert-Einstein-Institut),
             Am M\"uhlenberg 1, 14476 Potsdam, Germany}

\date{\today}

\begin{abstract}
Currently planned second-generation gravitational-wave laser
interferometers such as Advanced LIGO exploit the extensively
investigated signal-recycling (SR) technique. Candidate Advanced
LIGO configurations are usually designed to have two resonances
within the detection band, around which the sensitivity is
enhanced: a stable optical resonance and an unstable
optomechanical resonance -- which is upshifted from the pendulum
frequency due to the so-called {\it optical-spring} effect.
Alternative to a feedback control system, we propose an
all-optical stabilization scheme, in which a second optical spring
is employed, and the test mass is trapped by a stable
ponderomotive potential well induced by two carrier light fields
whose detunings have opposite signs. The double optical spring also brings additional flexibility in re-shaping the noise spectral density and optimizing toward specific gravitational-wave sources. The presented scheme can be extended easily to a multi-optical-spring system that allows further optimization.
\end{abstract}


\maketitle

\section{Introduction}

The array of the four large scale laser interferometers
(LIGO~\cite{SHOEMAKER2004}, VIRGO~\cite{FIORE2002},
GEO~\cite{WIL2002} and TAMA~\cite{ANDO2001}) represent the first
generation of interferometric gravitational-wave (GW) detectors ({\it interferometers} for short).
Next-generation interferometers, such as
Advanced LIGO~\cite{advLIGO}, plan to use the so-called {\it
detuned signal-recycling} (SR) technique --- in which an
 additional mirror is
placed behind the dark port of a Michelson
interferometer in order to modify the optical
resonance structure of the interferometer:
depending on the location and the reflectivity of this
signal-recycling mirror, the eigenfrequency and quality factor of
the optical resonance can be adjusted~\cite{Meers1988,VBMB1988,HSMSWWSRD98}.  As shown theoretically by
Buonanno and Chen~\cite{BUCH2001,BUCH2002,BUCH2003} and
experimentally by Somiya et al.~\cite{Som2005} and Miyakawa et
al.~\cite{Miyakawa2006}, detuned signal-recycling makes the power
inside the interferometer dependent on the motion of the mirrors,
creating an {\it optical spring}. With optical power as high as
planned for second-generation interferometers, detuned
signal-recycling interferometers are characterized by two
resonances within the detection band. One resonance is optical in
nature, while the other one is due to the optical spring: the
eigenfrequency of the test masses can be shifted from that of the
pendulum up into the detection band. The general principle
underlying the optical spring effect in signal-recycling interferometers
is identical to that explained by Braginsky, Gorodetsky
and Khalili~\cite{BSV1977} for a single detuned cavity (cf.~\cite{BUCH2003}),
which had been employed in their proposal of the
{\it optical bar} detection scheme~\cite{BGK1997}.

\begin{figure}[t]
\includegraphics[width=\linewidth]{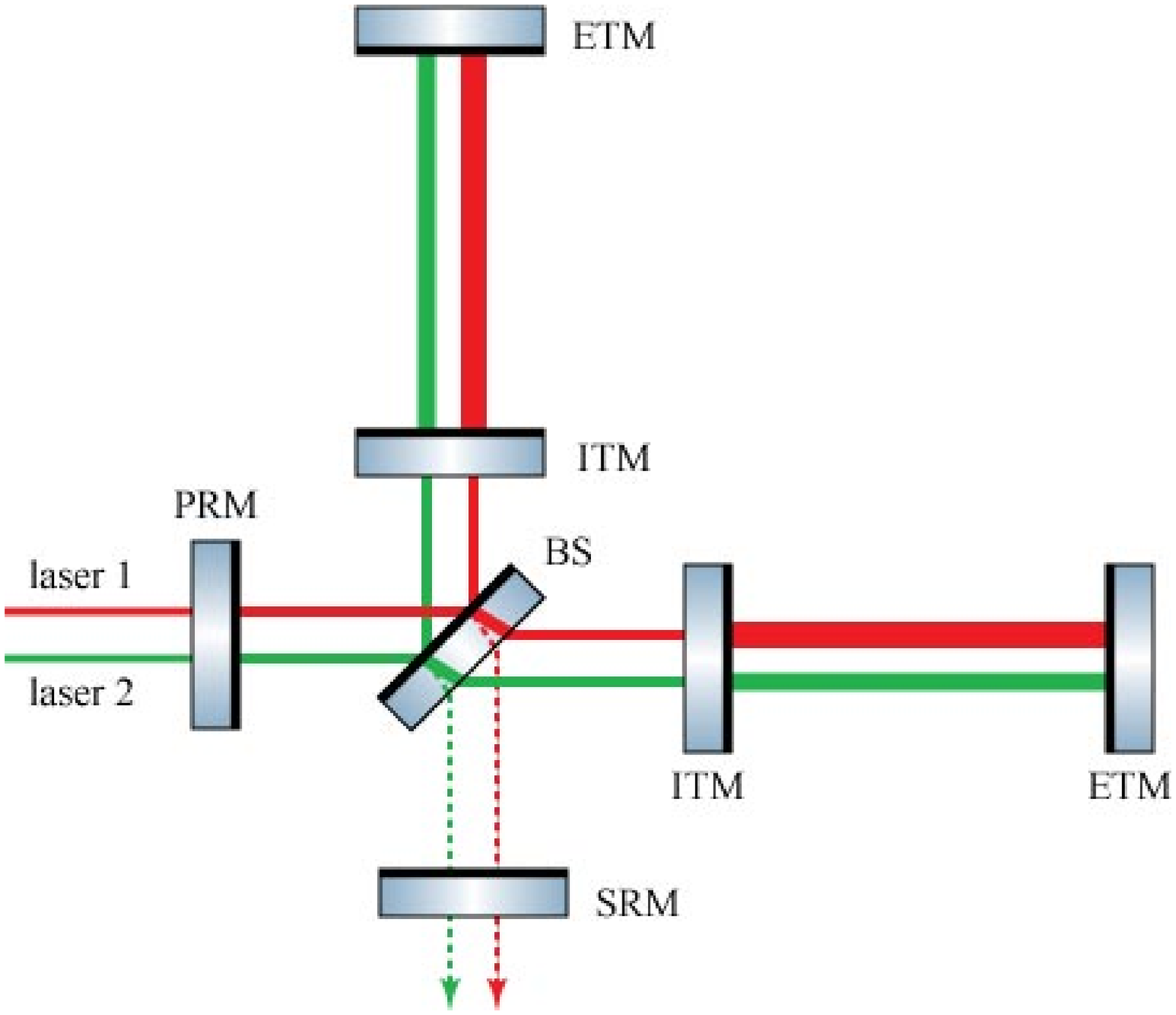}
\caption{Schematic plot of a power- and signal-recycled Michelson
interferometer with arm cavities and double optical spring, which
is realized by injecting two carriers with orthogonal polarizations. Both carriers are split at the beam splitter (BS) and transmitted
through the input mirrors (ITM) into the arm-cavities consisting
of ITM and the end mirrors (ETM); they both resonate
in the arm cavity. Two additional
cavities are realized by the signal-recycling mirror (SRM) and the power-recycling mirror
(PRM). } \label{Fig:dosmich}
\end{figure}

One concern that arises with the use of the optical spring is that
it always causes instability: depending on the sign of the
detuning, the optical force either brings anti-damping, or creates
an anti-spring. Buonanno and Chen have shown in
Ref.~\cite{BUCH2002} that one can cope with the instability by
incorporating a linear feedback control which ideally would not
modify the noise spectral density of the GW detector. In practice,
however, the need for control inside the detection band, can cause
undesirable complexity in the control system or additional
classical noise.

Here we propose an alternative way to suppress the instability, by
injecting a second carrier field from the bright port (cf.
Fig.\,\ref{Fig:dosmich}). We shall assume these carriers have
different polarizations (as in Ref.~\cite{RMSLSDC2007}), so that
there is no direct coupling between the two fields (although they
both directly couple to the mirrors). Differently from
Ref.~\cite{RMSLSDC2007}, in which the second carrier does not
enter the arm cavity at all, in this proposal, the second carrier
resonates in the arm cavity, but is subject to different SR
detuning and SR mirror reflectivity. Nevertheless,  similar to
Ref.~\cite{RMSLSDC2007}, a homodyne detection must be performed at
the dark port around each of the two carriers, with the two
outputs  combined with appropriate filters. The main purpose
of this second carrier is to create a second optical spring that
forms a stable optical spring together with the first one --- even
though each individual optical spring, acting alone, would be unstable.
Such a stable, Double Optical Spring (DOS) is possible at least in two ways. The first way ({\it
weak stabilization}) relies on the observation that the ratio
between the optical spring constant's real part (strength of
spring/anti-spring) and its imaginary part (strength of
damping/anti-damping) depends on the detuning frequency of the
carrier --- the detunings of the two carriers can be arranged such
that the first one has a stronger spring and a weaker
anti-damping, while the second one has a weaker anti-spring and a
stronger damping. In this case, stabilization can be
achieved with a weak second carrier, which does not modify
the sensitivity of the interferometer by much. The
second way ({\it annihilation}) requires the two carriers to have
equal power and exactly opposite detunings, such that their
optical springs exactly cancel each other, and the test-masses'
differential motion behaves as a free mass again. Interestingly,
this configuration gives exactly the noise spectrum that had been
expected by the GW community~\cite{GSSW1999} at the time an
interferometer with detuned signal-recycling was not thoroughly
studied, i.e. before Refs.~\cite{BUCH2001,BUCH2002,BUCH2003} were
published.

Although straightforward to understand, weak stabilization and
annihilation are by no means the only possibilities. In fact, an
additional benefit of the DOS technique is
that it increases the flexibility in shaping the noise curves: for
any specific source, the noise spectrum can be optimized
correspondingly over the parameter subspace of the two carriers,
subject to the constraint that the resulting dynamics must be
stable.  In this paper, we will carry out an optimization for
neutron-star-neutron-star binary inspirals --- using both the
current noise budget of Advanced LIGO and a plausible noise
budget for interferometers right beyond the second generation. Note that the parameters for the Advanced LIGO configuration as well as for the other single optical spring configuration have been obtained by using the same optimization.

The DOS technique can also be used for the stabilization of the
optical-spring ponderomotive squeezer, which generates frequency
independent squeezed vacuum below the optical spring
frequency~\cite{CCKOVWM2006,MRCWMSDC2008}. The stable
optomechanical resonance  has already been demonstrated
experimentally~\cite{CCIMORSWWM2007}.

This paper is organized as follows: In Sec.\,\ref{sec:dosgwd} we
shall motivate the stabilization process and studying the
classical dynamics of the double optical spring systems. In
Sec.\,\ref{sec:dos} we recall the necessary basics in order to
calculate the input-output relation of a DOS interferometer. In
Sec.\,\ref{sec:nsdq} different applications of the DOS are
discussed and our scheme is extended to a a multi-optical-spring
system. In Sec.\,\ref{sec:conclusion} we summarize our main
conclusions.

\section{Classical dynamics}
\label{sec:dosgwd}

In this section, we consider the classical dynamics of
double-optical-spring stabilization. For the
mechanical degrees of freedom, we only consider
the {\it differential mode}
\begin{equation}\label{Eq:diffmode}
x\equiv x_{\rm antisym}=(x_{\rm ETM}^{(n)}-
x_{\rm ITM}^{(n)})-(x_{\rm ETM}^{(e)}-x_{\rm ITM}^{(e)})
\end{equation}
between the interferometer's test-masses, which is sensed at the
dark port by both carriers.   Before coupling to the light, the differential
mode has an eigenfrequency of $\approx 1$\,Hz (that of the suspension system),
and effective mass of $m/4$, where $m$ is the mass of each individual mirror.
Since the pendulum frequency is far below the detection band, we will simply
treat  the mirrors as free masses.  For a carrier with angular frequency $\omega_0$ that is resonant in the arms, when the mirrors are held fixed, the optical
resonant frequency of the differential optical mode (to be precise, the one that is closest to this carrier) is given by $\omega_0 - \lambda-i\epsilon$, where, in terms of interferometer
parameters, $\epsilon$ and $\lambda$
are given by (cf.~\cite{BUCH2003})
\begin{eqnarray}
\lambda &=& \gamma_o \frac{2 \rho_{\rm SR} \sin(2
\phi)}{1 + (\rho_{\rm SR})^2 + 2
\rho_{\rm SR} \cos(2\phi)}\,, \label{Eq:lambda} \\
\epsilon &=& \gamma_o \frac{1 - (\rho_{\rm SR})^2}{1 + (\rho_{\rm
SR})^2 + 2 \rho_{\rm SR} \cos(2\phi)}\,.  \label{Eq:epsilon}
\end{eqnarray}
where $\gamma_0 = Tc/(4L)$ is the half line width of the arm cavity ($T$ the power transmissivity of the
ITM, $c$ the speed of light, and $L$ the arm length),
$\rho_{\rm SR}$ is the amplitude reflectivity of the SRM,
and $\phi$ the detuning phase of the carrier in the SR cavity (single trip).
In reality, when two carriers are both resonant in the arm cavity,
their detuning phases in the signal-recycling cavity must differ by

\begin{equation}\label{Eq:FSR}
\Delta \phi =\frac{2 \pi n l_{\rm SR}}{c}(\Delta \nu)_{\rm FSR},\quad n=0,\pm1,\pm2,\ldots
\end{equation}
where $(\Delta \nu)_{\rm FSR}$ is the free spectral range of the
arm cavities.  This constraint must be taken into account in practical designs of DOS interferometers.

Now with optomechanical coupling, let us first consider a single detuned
carrier. Treating the mirrors as free masses to start with, the
classical equation of motion of the differential mode can be
written in frequency domain as:
\begin{equation}\label{Eq:classdym}
- \frac{m}{4} \Omega^2 \ x (\Omega) =  - K_{\rm os}(\Omega) \ x
(\Omega) + F_{\rm ext}\,,
\end{equation}
where $m/4$ is the reduced
mass, $F_{\rm ext}$ is any external
classical force. The frequency dependent optical spring constant
is given by~\cite{BUCH2003}
\begin{equation} \label{Eq:Kpond}
K_{\rm os} = - \frac{ m  \theta}{4} \frac{\lambda}{(\Omega -
\lambda + {\rm
i}\epsilon) (\Omega + \lambda + {\rm i} \epsilon)}\,, \\
\end{equation}
where  $\theta$  is given by
 \begin{equation}
\theta = \frac{8 P \omega_0}{m L c}\,,
\end{equation}
with $P$ the carrier's circulating power in the arms. Note
that $\theta$ has units of ${\rm Hz}^3$.

Before treating a full-power interferometer, it is instructive to
first draw our attention to a {\it weakly coupled regime},
assuming that optical frequency scales (effective detuning
$\lambda$ and effective bandwidth $\epsilon$) are much larger than
the resulting optomechanical resonant frequency. In the weakly
coupled regime, one can expand the optical spring constant (cf.
e.g. Ref.~\cite{BUCH2002}) as
\begin{equation} \label{Eq:Kpondapprox}
K_{\rm os}(\Omega)\approx \frac{m\theta \lambda}{4
\left(\epsilon ^2+\lambda ^2\right)}\left(1+{\rm i}\frac{2\epsilon
\Omega }{\left(\epsilon ^2+\lambda ^2\right)}\right)\equiv K-{\rm
i}\Omega\Gamma \,.
\end{equation}
where $K$ and $\Gamma$ are both real constants. Analogous to a
mechanical spring, $K$ describes the restoring force while
$\Gamma$ denotes the damping. Real and imaginary parts of the
optical spring constant are proportional to $\theta$, which is in
turn proportional to the carrier power $P$. Inserting
Eq.\,\eqref{Eq:Kpondapprox} into Eq.\,\eqref{Eq:classdym},
stability requires
\begin{equation}\label{eqn:stablecondsimp}
K> 0 \quad \textrm{and} \quad \Gamma>0\,,
\end{equation}
or basically positive spring constant and positive damping.
\begin{figure}[h]
\includegraphics[width=6.5cm,angle=-90]{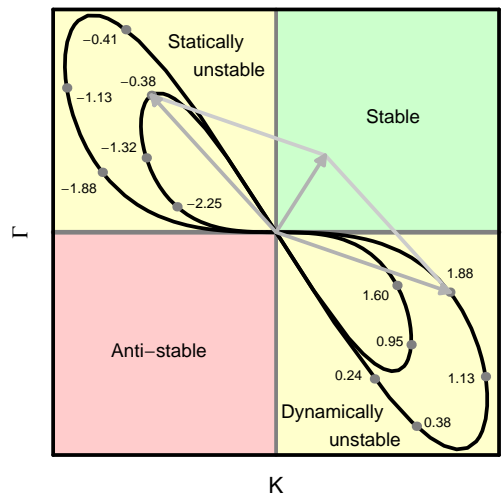}
\caption{Real and imaginary parts of the spring constant $K_{\rm
os}$ in the weakly coupled regime [Eq.~\eqref{Eq:Kpondapprox}]. For each trajectory the circulating power and bandwidth are fixed to a certain value while detuning varies from $-\infty$ to $\infty$.  Outer trajectory corresponds to
a higher circulating power. Example values of
$\lambda/\epsilon$ are marked on the trajectories.}
\label{Fig:spring}
\end{figure}

From Eq.~\eqref{Eq:Kpondapprox}, it is straightforward
to deduce that the stability condition~\eqref{eqn:stablecondsimp} can never be fulfilled
by a single optical system, since a positive detuning produces
always anti-damping ($\Gamma<0$) while a negative detuning always
comes along with an anti-restoring force ($K<0$). In
Fig.\,\ref{Fig:spring}, for fixed circulating power and effective
line width ($\epsilon$), as the effective detuning ($\lambda$)
shifts from $-\infty$ to $\infty$, we plot the trajectory mapped out
by $(K,\,\Gamma)$. We use two different powers, with outer trajectory corresponding
to a higher power. Indeed, these  $\infty$-shaped trajectories are
confined within quadrants with $K \cdot \Gamma <0$. In case of a
double optical spring, each individual spring, as we change its
detuning frequency,  has its own $\infty$-shaped trajectory. When
two optical springs combine, their complex spring constants add
up, which can be  depicted by a vector addition in
Fig.\,\ref{Fig:spring}. By adjusting the detunings for first and
second carrier it is possible to find many stable compositions,
with one of them depicted in Fig.~\ref{Fig:spring}. In
this configuration, a relatively strong optical spring
is stabilized by a relatively weak anti-spring generated by a lower power.  This
is possible because the stronger optical spring is generated by a carrier with
relatively high optical quality factor, $|\lambda/\epsilon|$, which
tends to yield a stronger restoring (or anti-restoring) than damping (or anti-damping),
while the weak anti-spring is generated by a carrier with low optical quality factor,
which tends to yield a stronger damping (or anti-damping) than restoring (or anti-restoring).  A lower optical power of the second carrier allows the damping of the second spring to match that of the first one, while makes the anti-restoring of the second spring much weaker than the restoring of the first spring. Mathematically, this {\it weak stabilization} can be summarized as
\begin{equation}
\frac{|K^{(1)}|}{|\Gamma^{(1)}|} \gg \frac{|K^{(2)}|}{|\Gamma^{(2)}|},\,
|\Gamma^{(1)}| \sim |\Gamma^{(2)}| \Rightarrow |K^{(1)}| \gg |K^{(2)}|.
\end{equation}
As another (rather extreme) example of DOS stabilization, we note that when
the two carriers have the same power and bandwidth, but
the opposite detuning, their optical spring constants
exactly cancel with each other.  In fact, this cancelation, or annihilation, is valid for an arbitrarily
strong coupling, cf.~Eq.~\eqref{Eq:Kpond}. Mathematically, this can be summarized as
\begin{equation}
K_{\rm os}(\theta,\epsilon,\lambda;\Omega) +K_{\rm os}(\theta,\epsilon,-\lambda;\Omega)=0.
\end{equation}

As stated before, the stability condition given in
Eq.~(\ref{eqn:stablecondsimp}) is an approximation only valid in
the weakly coupled regime. A more precise statement regarding the
stability of the two carrier system is given by the condition that
all roots of the characteristic equation
\begin{equation}\label{eqn:stablecond}
-\frac{m}{4}\Omega^2 + K_{\rm os}^{(1)}(\Omega)+ K_{\rm
os}^{(2)}(\Omega)=0
\end{equation}
must have negative imaginary parts.
\begin{figure}[h!!!]
\includegraphics[height=\linewidth,angle=-90]{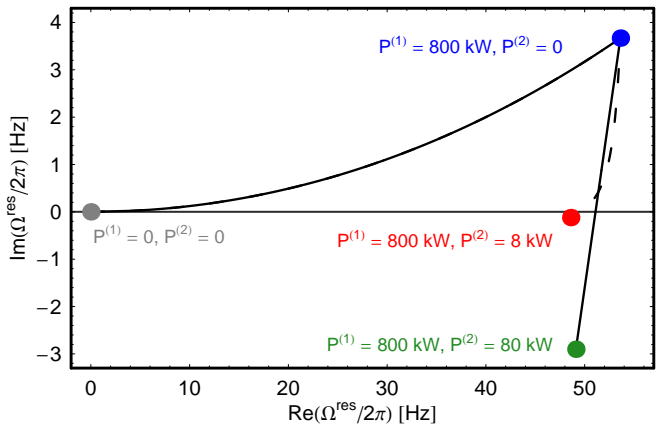}
\caption{Example of DOS stabilization process, as illustrated
by trajectories of the optomechanical resonant frequency in the
complex plane [Eq.~\eqref{eqn:stablecond}].
The test masses start off as free masses (gray dot).
As the first carrier light ($\epsilon^{(1)}=2\pi\ 120
{\rm Hz}$ and $\lambda^{(1)}=2\pi\ 290 {\rm Hz}$) is turned on,
it causes an upshift of the mechanical resonant frequency , as well as  a mild
anti-damping.  The trajectory ends at the blue dot, with a circulating
power of 800\,W.   Subsequently, the
second carrier is also turned on, bringing a damping while slightly downshifting the optomechanical resonant frequency. For $\epsilon^{(2)}=2\pi\ 5 {\rm Hz}$,
$\lambda^{(2)}=-2\pi\ 55 {\rm Hz}$, the trajectory ends at the
 red dot, with a circulating power of $8 \ {\rm kW}$;  while for $\epsilon^{(2)}=2\pi\ 60 {\rm Hz}$, $\lambda^{(2)}=-2\pi\ 60 {\rm
Hz}$ the trajectory ends at the green dot, with circulating power of $80 \ {\rm kW}$. \label{Fig:stab}}
\end{figure}
In Fig.~\ref{Fig:stab}, we explore high-power DOS stabilization by
tracing the real and imaginary parts of the optomechanical
eigenfrequency, obtained numerically solving
Eq.\,\eqref{eqn:stablecond}. We first consider
single-optical-spring configurations with power increasing from 0
to 800\,KW (from left most dot to top dot). In this case both the real part of the optomechanical
resonant frequency and the anti-damping increases. Then we fix
the first carrier at 800\,kW and increase the second carrier (up to 8\,kW
and 80\,kW, respectively, for two different choices of linewidth and
detuning),
which stabilizes the system by adding damping while only slightly
decreases the optomechanical resonant frequency by a few Hz (from the top dot to the two lower-right dots). Here we do not plot the optical resonances, which for the circulating power considered here remain stable.

Before ending this section, we note that, when pendulum frequency
is not neglected, there does exist a stable single optical spring
regime where an increase in mechanical resonant frequency is
associated with an increase in damping. But this requires that the
optical frequency scales be lower than the pendulum frequency,
which is not desirable in case of GW detectors. Such a regime was
experimentally investigated by Schliesser et al.~\cite{SDNVK2006}.

\section{Input-output relation and combined noise spectral density}\label{sec:dos}

In this section, we consider the sensitivity of stable DOS
interferometers.  Since the two carrier fields
are different in polarization and frequency, both fields can be
measured independently and we effectively obtain two
interferometers in one scheme, both sensing the same differential mode [Eq.~\eqref{Eq:diffmode}] of
mirror motion. The input optical vacuum fluctuations associated with the two carriers, on the other hand, are independent.

Throughout this paper, we will assume that GWs with amplitude $h$
incident from right above the detector plane, with a polarization
that maximizes the response of our $L$-shaped Michelson
interferometer. In the following we will list the Heisenberg
equations of motion in frequency domain
\cite{KLMTV2001,BUCH2001,BUCH2002,BUCH2003,CCM2005} for the
antisymmetric mode of motion of the arm cavity mirrors $\hat x$
and for the two measurement outputs $\hat y^{(i)}$:
\begin{align}
\hat x =& - R_{xx}(\Omega)[\hat F^{(1)} (\Omega) + \hat F^{(2)}
(\Omega)\notag\\
& + (R_{FF}^{(1)} (\Omega)+R_{FF}^{(2)} (\Omega)) \ \hat x] + L\
h + \hat \xi_{\rm noise}
\label{eomx}\\
\hat y^{(1)} =& \hat Y_1^{(1)} (\Omega) \ \sin \zeta^{(1)} + \hat
Y_2^{(1)} (\Omega) \ \cos \zeta^{(1)}\notag\\
& + [R_{Y_1F}^{(1)} (\Omega) \ \sin \zeta^{(1)} + R_{Y_2F}^{(1)}
(\Omega) \ \cos \zeta^{(1)}] \hat x
\,, \label{eom1} \\
\hat y^{(2)} =& \hat Y_1^{(2)} (\Omega) \ \sin \zeta^{(2)} + \hat
Y_2^{(2)} (\Omega) \ \cos \zeta^{(2)}\notag\\
& + [R_{Y_1F}^{(2)} (\Omega) \ \sin \zeta^{(2)} + R_{Y_2F}^{(2)}
(\Omega) \ \cos \zeta^{(2)}]\hat x\,. \label{eom2}
\end{align}
Note that we have labeled all quantities with superscripts $(1)$
and $(2)$ for the first carrier and the second carrier,
respectively. It was shown in Ref.~\cite{HSD2004} that the
interferometer's output is only marginally influenced by seismic
noise, thermal noise and radiation pressure noise introduced at
the beam splitter, since the carrier light incident on the beam
splitter is weak and the arm cavities prevent fluctuations from
building up. The out-going sideband fields at the dark port around
the two different carrier fields are split and each is sensed
independently by a homodyne detection scheme, which measures a
certain combination of amplitude and phase quadratures (described
by the phases $\zeta^{(1),(2)}$). The operator $\hat F^{(i)}$ in
Eq.\,(\ref{eomx}) describes the radiation pressure forces which
would act on fixed mirrors due to the incoming vacuum fields at
the dark port. The operators $\hat Y^{(i)}_j$ in
Eqs.\,(\ref{eom1}) and (\ref{eom2}) account for the out-going
fluctuations in the quadratures in case of fixed mirrors.
(In language of Ref.~\cite{BUCH2003}, $\hat F^{(i)}$ and $\hat
Y^{(i)}_j$ are {\it free quantities}). The
operators $\hat \xi_{\rm noise}$ describe the classical displacement
noise of the differential mode. The quantity
$R_{xx} \equiv -4/(m
\Omega^2)$ is the mechanical susceptibility of the differential mode, the susceptibilities $R^{(i)}_{FF} = - K_{\rm
os}^{(i)}$ [cf.~Eq.~\eqref{Eq:Kpond}] correspond to the optical spring constants,  and
$R^{(i)}_{Y_{i}F}$ are optical transfer functions from the differential
mode to the out-going quadrature fields.

According to Ref.~\cite{BUCH2003}, the free quantities $\hat F^{(i)}$
and $\hat Y_j^{(i)}$ are related to incoming amplitude and phase quadratures, $\hat a_1^{(i)}$ and $\hat a_2^{(i)}$, as
\begin{align}
\hat F^{(i)} &= \sqrt{\frac{\epsilon^{(i)} \theta^{(i)}m \hbar
}{2}} \frac{({\rm i} \Omega - \epsilon^{(i)}) \ \hat a_1^{(i)} +
\lambda^{(i)} \ \hat a_2^{(i)}}{(\Omega+{\rm i}\epsilon^{(i)})^2-(\lambda^{(i)})^2}, \\
\hat Y_1^{(i)} &= \frac{[(\lambda^{(i)})^2 - (\epsilon^{(i)})^2 -
\Omega^2] \ \hat a_1^{(i)} + 2 \lambda^{(i)} \epsilon^{(i)} \ \hat
a_2^{(i)}}{(\Omega+{\rm i}\epsilon^{(i)})^2-(\lambda^{(i)})^2}, \\
\hat Y_2^{(i)} &= \frac{\big[(\lambda^{(i)})^2 - (\epsilon^{(i)})^2 - \Omega^2\big]
\hat a_2^{(i)}-2 \lambda^{(i)} \epsilon^{(i)} \ \hat
a_1^{(i)} }{(\Omega+{\rm i}\epsilon^{(i)})^2-(\lambda^{(i)})^2}.
\end{align}
For vacuum fluctuations, we have
\begin{equation}
\langle \hat a_k^{(i)} (\Omega)\ (\hat a_l^{(j)})^\dagger
(\Omega') \rangle_{{\rm sym}} = \pi \ \delta (\Omega - \Omega') \
\delta_{ij} \ \delta_{kl}\,.
\end{equation}
The optical transfer functions are given by~\cite{BUCH2003}
\begin{align}
R_{Y_1F}^{(i)} &= \sqrt{\frac{\epsilon^{(i)} \theta^{(i)} m}{2
\hbar}} \frac{\lambda^{(i)}}{(\Omega+{\rm i}\epsilon^{(i)})^2-(\lambda^{(i)})^2}\,,  \\
R_{Y_2F}^{(i)} &= - \sqrt{\frac{\epsilon^{(i)} \theta^{(i)} m}{2
\hbar}} \frac{\epsilon^{(i)} - {\rm i} \Omega}{(\Omega+{\rm i}\epsilon^{(i)})^2-(\lambda^{(i)})^2}\,,
\end{align}
Finally, the classical noise operator $\hat \xi_{\rm noise}$ satisfies
\begin{equation}
\langle \hat \xi_{\rm noise} (\Omega)\ \hat
\xi_{\rm noise}^\dagger (\Omega') \rangle_{{\rm sym}} = \pi \ \delta
(\Omega - \Omega') \ L^2 S_h^{\rm cl} (\Omega)\,,
\end{equation}
where $S_h^{\rm cl}$ is the classical noise in the gravitational-wave strain.

\begin{figure}[t!]
\includegraphics[width=6.5cm,angle=0]{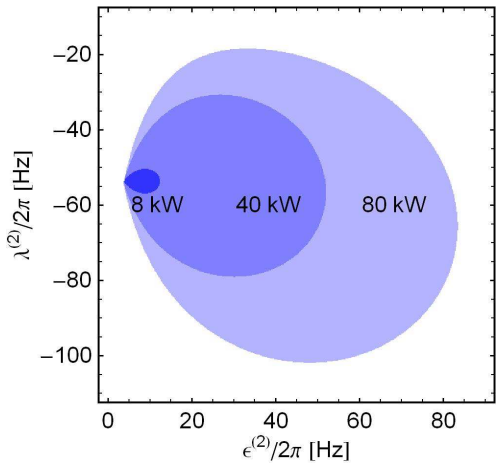}
\caption{Allowable regions for the optical resonances of the second carrier,
given a circulating power of 8\,kW, 40\,kW and 80\,kW,
in order to stabilize a 800\,kW first carrier that corresponds to the Advanced LIGO narrowband
mode (cf. Tab. \ref{Tab:definitions}).} \label{Fig:stableregion}
\end{figure}

The outgoing sideband fields at the dark port around the two
different carriers are detected independently via homodyne detection. We seek a linear
combination of the two measurement output channels, $\hat y^{(1)}$
and $\hat y^{(2)}$ given in Eqs.\,(\ref{eom1}) and (\ref{eom2}),
\begin{equation}\label{combiout}
\hat y = K_1 (\Omega) \ \hat y^{(1)} + K_2 (\Omega) \ \hat
y^{(2)}\,,
\end{equation}
which has optimal sensitivity to the GW strain $h$. Namely, one
has to minimize the $h$-referred noise spectral density by varying the two
filter functions $K_i (\Omega)$. The optimal solution can be found by a straightforward calculation described in
detail for a general multi-channel interferometer in
App.~\ref{sec:mos}.

\section{Example configurations}\label{sec:nsdq}
In this section, different example configurations of a double
optical spring interferometer are discussed. At first the
considerations are restricted to quantum noise only in order to
clarify two distinct regimes of our proposed scheme. Subsequently
a configuration with realistic classical noise budget is
investigated and optimized for neutron-star-neutron-star
binary inspirals. Finally, we lower the
classical noise budget and explore whether DOS
configurations can take full advantage of this improvement.
We are aiming at applying the DOS configuration as
an upgrade candidate for the Advanced LIGO detector.

\subsection{Quantum noise examples}
Here we study the quantum noise spectrum of two special
regimes of DOS: weak stabilization and annihilation.
\begin{table*}
\begin{center}
\begin{tabular}{cccc}
\hline\hline
  Symbol & physical meaning & AdvLIGO narrowband & AdvLIGO broadband \\
  \hline
  $m$ & single mirror mass & $40 \ {\rm kg}$ & $40 \ {\rm kg}$ \\
  $2\pi c/\omega_0^{(1)}$ & laser wavelength & $1064 \ {\rm nm}$ & $1064 \ {\rm nm}$ \\
  $P^{(1)}$ & circulating power & $800 \ {\rm kW}$ & $800 \ {\rm kW}$\\
  $L^{(1)}$ & interferometer arm length  & $4 \ {\rm km}$ & $4 \ {\rm km}$\\
  $\phi^{(1)}$ & detuning phase of SR cavity & $2\pi \ 0.242$ & $2\pi \ 0.247 $\\
  $\rho_{\rm SR}$ & signal-recycling mirror reflectivity & $\sqrt{0.93}$ & $\sqrt{0.93}$\\
  $\gamma_o$ & cavity half bandwidth  & $2 \pi \ 15 \ {\rm Hz}$ & $2 \pi \ 15 \ {\rm Hz}$\\
  $\zeta^{(1)}$ & detection angle & $2 \pi \ 0.347$ & $2\pi \ 0.45$\\
  $\epsilon$ & effective half bandwidth & $2\pi \ 120 \ {\rm Hz}$ & $2 \pi \ 395 \ {\rm Hz}$ \\
  $\lambda$ & effective detuning & $2\pi \ 290 \ {\rm Hz}$ & $2\pi \ 411 \ {\rm Hz}$
  \\
\hline\hline
\end{tabular}
\caption{Parameter values for  single optical spring
Advanced LIGO interferometer configurations used throughout the
calculations. The narrowband configuration is optimized for NS-NS
binaries using the current Advanced LIGO noise budget. For the
broadband operational mode, we allow 10\% decrease in detectable
distance for NS-NS binaries assuming the current Advanced LIGO
noise budget, while maximizing the contribution to SNR from frequencies above 500\,Hz}.\label{Tab:definitions}
\end{center}
\end{table*}

{\bf Weak stabilization.} In this scenario, we use
a relatively weak second carrier to stabilize a typical Advanced LIGO configuration.  In Fig.~\ref{Fig:stab}, examples were already given, in which a second carrier of 8\,kW and 80\,kW, respectively, has
been used to stabilize the narrowband mode of Advanced LIGO (cf.~Tab.~\ref{Tab:definitions}), while Fig.~\ref{Fig:stableregion} provides the region
of all possible second carriers, given circulating power of 8\,kW, 40\,kW and 80\,kW.

\begin{figure}[t!]
\includegraphics[height=\linewidth,angle=-90]{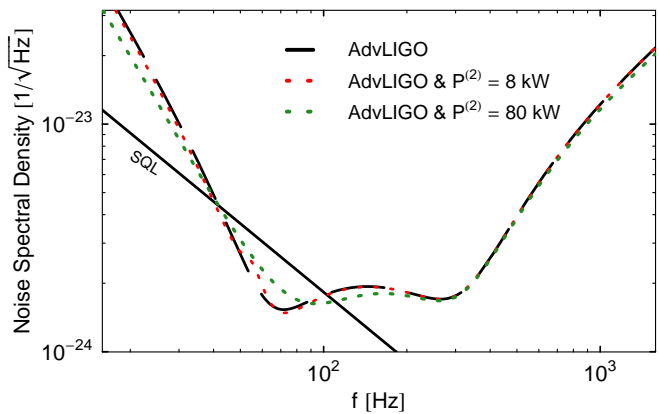}
\caption{Noise spectrum of weak-stabilization DOS configurations,
with parameters corresponding to configuration in Fig.~\ref{Fig:stab}.
Namely, the first carrier is identical to the carrier in the Advanced LIGO
narrowband mode (cf. Tab.
\ref{Tab:definitions}). Two possible choices for the stabilizing
second carrier are: (i) $P^{(2)}=8 \ {\rm kW}$,
$\epsilon^{(2)}=2\pi\ 5 {\rm Hz}$, $\lambda^{(2)}=-2\pi\ 55 {\rm
Hz}$ and (ii) $P^{(2)}=80 \ {\rm kW}$, $\epsilon^{(2)}=2\pi\ 60 {\rm
Hz}$, $\lambda^{(2)}=-2\pi\ 60 {\rm Hz}$. Phase quadrature detection ($\zeta^{(2)}=0$) in both cases.}\label{Fig:NSD4}
\end{figure}

In Fig.~\ref{Fig:NSD4}, we plot the noise spectrum of DOS
interferometers that correspond to the two stabilizing configurations
in Fig.~\ref{Fig:stab}, namely with first carrier identical to
the carrier of the Advanced LIGO narrowband mode, with
resonant frequencies and powers of second carrier given in
Fig.~\ref{Fig:stab}, and a phase readout quadrature associated with the second carrier (i.e., $\zeta^{(2)}=0$). As we see from Fig.~\ref{Fig:NSD4}, the DOS noise spectra in both cases do not differ much from that of the Advanced LIGO narrowband mode.

\begin{figure}[h!!!]
\includegraphics[height=\linewidth,angle=-90]{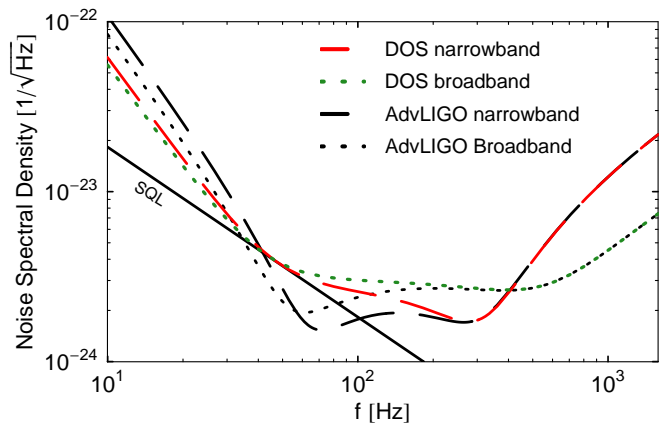}
\caption{Noise spectrum of DOS configurations with
canceled optical springs, based on Advanced LIGO
narrowband and broadband modes
(cf. Tab. \ref{Tab:definitions}). Detuning and detection
angle of the first carrier are identical to that of Advanced LIGO
narroband mode, while those of the second are opposite. The total power is equally
divided into two parts ($P^{(1)}=P^{(2)}=400\ {\rm kW}$). Other
parameters are left unchanged.}\label{Fig:NSD3}
\end{figure}

{\bf Annihilation.} Now we turn to a different situation where
both carriers have half the Advanced LIGO circulating power,
namely $400 \ {\rm kW}$ with the two detunings exactly opposite.
In this case the two optical springs cancel each other and the total
effective ponderomotive rigidity vanishes. For Fig.\,\ref{Fig:NSD3} the two detection angles also have opposite signs while the absolute
values of the two detunings (detection angles) agree with the
detuning (detection angle) in case of the conventional Advanced
LIGO configurations (cf. Tab.\,\ref{Tab:definitions}). All other
parameters are left unchanged.
\begin{figure}[h!!!]
\includegraphics[height=\linewidth,angle=-90]{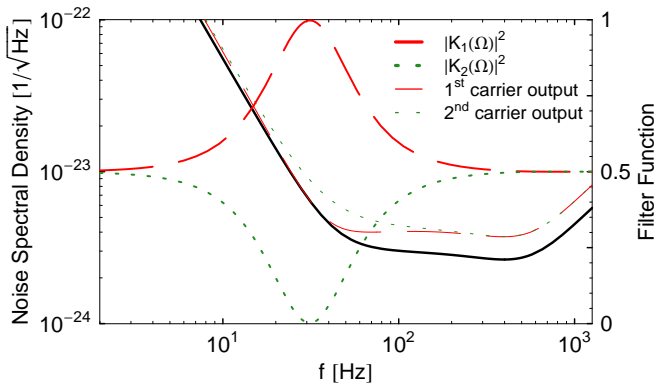}
\caption{Example by means of the broadband configuration (cf. Fig.\,\ref{Fig:NSD3}): Squared absolute values of filter functions for the two outputs. These curves actually account for how much of the contributions is used -- but they cannot show how the correlations among the two outputs have influence on the total output. The contributions correspond to measuring one of the outputs alone without filtering -- but still in presence of both carriers.
}\label{Fig:Filter}
\end{figure}

The sensitivity does not change in the high frequency regime where
shot noise is the limiting factor (cf. Fig.\,\ref{Fig:NSD3}). This
is because, generally, the noise spectral density of the shot noise remains unchanged under reversing the sign of the detuning and the
detection angle. With the optimal filter (cf.
Fig.\,\ref{Fig:Filter}), we obtain the same shot noise as the
Advanced LIGO configuration, as the total power is conserved. In the low frequency regime the sensitivity is slightly
improved compared to the single optical spring interferometer. It
is well-known that the quantum-noise limited sensitivity of single
optical spring interferometers at frequencies below the
optomechanical resonance is dramatically lower than the
sensitivity of non-optical-spring interferometers because the
strong restoring force due to the single optical spring suppresses
the response of the interferometer's differential mode to GWs. In
a double optical spring interferometer the second carrier usually
results in a less rigid or completely canceled effective optical
spring. This effect appears in the first example where only a weak
second carrier is used (cf. Fig.\,\ref{Fig:NSD4}) and becomes more
significant here in the case of canceled optical springs (cf.
Fig.\,\ref{Fig:NSD3}). For intermediate frequencies the
sensitivity becomes worse due to the absence of the optomechanical
resonance gain. In this regime the output associated with the carrier having positive detuning usually has more sensitivity than the other one. Therefore, as shown in Fig.\,\ref{Fig:Filter}, the filters make sure that only this output contributes to the total output. Note that when the two outputs are filtered appropriately, the total noise curve can in fact be below the single contributions.

Since the optomechanical resonance disappears completely, the
noise spectral density in this configuration is equal to
\begin{equation}
S_h(\Omega)=S_{\rm shot}(\Omega)+\frac{S_{\rm SQL}^2(\Omega)}{4 S_{\rm shot}(\Omega)}\,.
\end{equation}
Here $S_{\rm shot}(\Omega)$ denotes the shot-noise spectral
density of a detuned single carrier interferometer with the same
total power as in the canceled optical spring configuration. The
free-mass standard quantum limit (SQL) \cite{BrKh1999a} for
detecting the gravitational-wave strain $h$ with a Michelson
interferometer with arm-cavities is given by
\begin{equation}
S_{\rm SQL}(\Omega)=\sqrt{\frac{8\hbar}{m\Omega^2 L^2}}\,.
\end{equation}

\begin{table*}
\begin{center}
\begin{tabular}{cccccccccc}
\hline\hline
  & $P^{(1)}$ in kW & $P^{(2)}$ in kW & $\lambda^{(1)}$ in Hz & $\lambda^{(2)}$ in Hz & $\epsilon^{(1)}$ in Hz & $\epsilon^{(2)}$ in Hz & $\zeta^{(1)}$ in radian & $\zeta^{(2)}$ in radian& Improvement \\
  \hline \multirow{7}{*}{\begin{sideways}{\centering high classical noise}\end{sideways}}
  & $800$ & $0$ & $2\pi \ 290$ & - & $2\pi \ 120$ & - & $2\pi \ 0.347$ & - & - \\
  & $750$ & $50$ & $2\pi \ 190$ & $-2\pi \ 70$ & $2\pi \ 50$ & $2\pi \ 20$ & $2\pi \ 0.125$ & $2\pi \ 0.075$ & 7.3\% \\
  & $700$ & $100$ & $2\pi \ 190$ & $-2\pi \ 50$ & $2\pi \ 40$ & $2\pi \ 30$ & $2\pi \ 0.0625$ & $2\pi \ 0.475$ & 22.6\% \\
  & $600$ & $200$ & $2\pi \ 180$ & $-2\pi \ 40$ & $2\pi \ 30$ & $2\pi \ 50$ & $2\pi \ 0.0125$ & $2\pi \ 0.3875$ & 34.4\% \\
  & $500$ & $300$ & $2\pi \ 150$ & $-2\pi \ 50$ & $2\pi \ 45$ & $2\pi \ 60$ & $2\pi \ 0.0125$ & $2\pi \ 0.375$ & 36.3\% \\
  & $450$ & $350$ & $2\pi \ 160$ & $-2\pi \ 10$ & $2\pi \ 30$ & $2\pi \ 40$ & $2\pi \ 0$ & $2\pi \ 0.3$ & 35.5\% \\
  & $400$ & $400$ & $2\pi \ 140$ & $-2\pi \ 20$ & $2\pi \ 20$ & $2\pi \ 55$ & $2\pi \ 0.01$ & $2\pi \ 0.31$ & 35.5\% \\
  \hline \multirow{6}{*}{\begin{sideways}{low classical noise}\end{sideways}}
  & $800$ & $0$ & $2\pi \ 170$ & - & $2\pi \ 10$ & - & $2\pi \ 0.263$ & - & - \\
  & $700$ & $100$ & $2\pi \ 165$ & $-2\pi \ 75$ & $2\pi \ 5$ & $2\pi \ 25$ & $2\pi \ 0.48$ & $2\pi \ 0.02$ & 32.1\% \\
  & $600$ & $200$ & $2\pi \ 150$ & $-2\pi \ 20$ & $2\pi \ 5$ & $2\pi \ 45$ & $2\pi \ 0.01$ & $2\pi \ 0.36$ & 52.2\% \\
  & $500$ & $300$ & $2\pi \ 140$ & $-2\pi \ 20$ & $2\pi \ 5$ & $2\pi \ 45$ & $2\pi \ 0.03$ & $2\pi \ 0.33$ & 74.3\% \\
  & $450$ & $350$ & $2\pi \ 135$ & $-2\pi \ 25$ & $2\pi \ 5$ & $2\pi \ 50$ & $2\pi \ 0.01$ & $2\pi \ 0.33$ & 83.4\% \\
  & $400$ & $400$ & $2\pi \ 127$ & $-2\pi \ 11$ & $2\pi \ 4$ & $2\pi \ 46$ & $2\pi \ 0.01$ & $2\pi \ 0.3$ & 110\% \\
\hline\hline
\end{tabular}
\caption{Parameters for double optical spring scheme optimized for
NS-NS binary systems. The total power is fixed to $800 \ {\rm
kW}$. The last column gives the improvement in event rate for our
proposed scheme compared to an optimized Advanced LIGO
configuration provided in the first row. For the upper part we adopted the
current Advanced LIGO noise budget and for the lower part the gravity gradient noise
and the suspension thermal noise are reduced by a factor of ten and the coating thermal
by a factor of three.
}\label{Tab:improvement}
\end{center}
\end{table*}

\subsection{Optimized configurations with Advanced LIGO classical noise buget}\label{sec:nsdcl}

In the following we optimize the double optical spring
interferometer for neutron star - neutron star (NS-NS) binary
inspirals. For such systems the last stable circular orbit gives
an upper frequency limit of $f_{\rm max}\approx 1570\,{\rm Hz}$
and seismic noise defines a lower bound of $f_{\rm min}\approx
7\,{\rm Hz}$, and the signal-to-noise ratio (SNR) in power
achievable by optimal filtering, for
a source at a given radius,  is given by
\begin{equation}\label{Eq:SNR}
\mathrm{SNR}^2 \propto \int_{f_{\rm min}}^{f_{\rm
max}}\frac{f^{-7/3}}{S_h(f)} \ {\rm d}f\,.
\end{equation}
Here $S_h(f)$ denotes the single-sided noise spectral
density of the interferometer. Note that this optimization
strategy tend to focus more on the low frequency regime at the
expense of the sensitivity at higher frequencies -- due to the
rather steep power law of $f^{-7/3}$. Since the radius of
detectable range is proportional to the SNR at a fixed radius, and that
the event rate is roughly the cube of the radius of detectable range,
the event rate  is proportional
to $\mathrm{SNR}^3$, or
\begin{equation}\label{Eq:ER}
\mbox{Event Rate} \propto \left[\int_{f_{\rm min}}^{f_{\rm
max}}\frac{f^{-7/3}}{S_h(f)} \ {\rm d}f\right]^{3/2}\,.
\end{equation}

\begin{figure}[t]
\includegraphics[height=\linewidth,angle=-90]{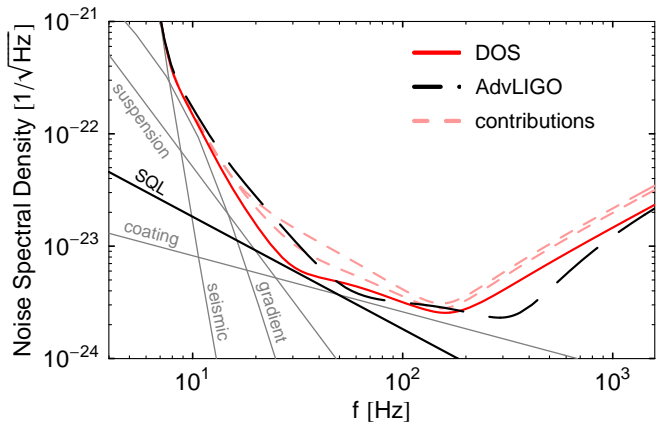}
\caption{Comparison of single optical spring and double carrier
with canceled optical spring Advanced LIGO configuration: the
same reflectivity of the SR mirror is assumed for both carriers.
We obtain 15\% improvement in event rate for DOS interferometer
using the following parameters: $P^{(1)}=P^{(2)}=400 \ {\rm kW}$,
$\rho_{\rm SRM}^{(1)}=\rho_{\rm SRM}^{(2)}=0.87$,
$\phi^{(1)}=-\phi^{(2)}=2 \pi \ 0.233$, $\zeta^{(1)}=2\pi \ 0.433 $
and $\zeta^{(2)}=0$.}\label{Fig:NSD12}
\end{figure}

\begin{figure}[t]
\includegraphics[height=\linewidth,angle=-90]{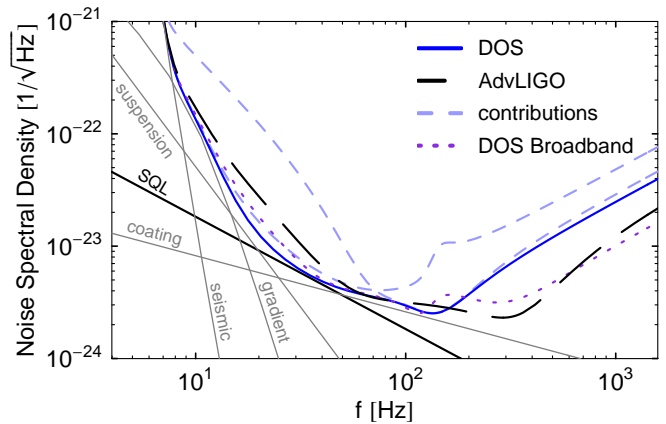}
\caption{Comparison of single and double optical spring Advanced
LIGO configuration: different SR mirror reflectivities for first
and second carrier are allowed. Parameters are given in detail in
last row of first block in Tab.~\ref{Tab:improvement} for the
narrowband configuration. For the broadband we have:
$\lambda^{(1)} = 2\pi \ 130 \ {\rm Hz}$, $\epsilon^{(1)} = 2\pi \
5 \ {\rm Hz}$, $\zeta^{(1)} = 2\pi \ 0.01$, $\lambda^{(2)} = -2\pi
\ 230 \ {\rm Hz}$, $\epsilon^{(2)} = 2\pi \ 155 \ {\rm Hz}$ and
$\zeta^{(2)} = 2\pi \ 0.02$. We obtain 35.5\% improvement in event
rate in case of the narrowband configuration and 16.4\% for the
broadband configuration.}\label{Fig:NSD22}
\end{figure}

For the optimization we have taken the current Advanced LIGO
classical noise budget into account (as given in \emph{Bench}
\cite{bench}): each contribution to the total classical noise
budget, i.e. suspension thermal noise, seismic noise, thermal
fluctuations in the coating and gravity gradient noise, are
presented in Figs.\,\ref{Fig:NSD12}--\ref{Fig:NSD22}.

It turns out that the second carrier not only helps to stabilize
the interferometer but can also improve its sensitivity. We first
assume both carriers to have the same SR mirror reflectivity, in
this case the gain in NS-NS sensitivity is maximized when the two
optical springs totally cancel each other (Fig.\,\ref{Fig:NSD12})
at an equal power distribution in the two carrier fields.
Fig.\,\ref{Fig:NSD12} also shows the contribution of each carrier
to the total noise spectral density.

If each carrier senses a different reflectivity of the SR mirror
the interferometer's sensitivity can be improved further. With
this additional degree of freedom the situation of canceled
optical springs is no longer the optimal choice for NS-NS
binaries. Each carrier can be optimized for a different frequency
regime such that they complement each other. This is illustrated
for an equal distribution of the total power ($400\,{\rm k}W$ for
each carrier) in Fig.\,\ref{Fig:NSD22} where one carrier ensures
good sensitivity in the low frequency regime while the other one
gives the main contribution at frequencies above the
optomechanical resonance as shown by the contribution curves. The
noise spectral density is close to the classical noise level in low frequencies, and an improvement in event rate of 35.5\% can be achieved for NS-NS
binaries for which the optimization was performed. As a
more general optimization is performed, allowing two carriers
to have different powers summing up to 800\,kW, it turns out
that $P^{(1)}=500\ {\rm kW}, \ P^{(2)}=300\
{\rm kW}$ achieves a slightly better event rate improvement of
36.3\%.  Results for different power distributions are given in
Tab.\,\ref{Tab:improvement}.


The sensitivity of all optimized configurations is basically
improved at low frequencies at the expense of the high
frequencies. Therefore it might be necessary to carry out an
additional broadband optimization in order to achieve a better
sensitivity in the high frequency regime. This can be accomplished
by first picking out all configurations obeying an event rate
which is at least a certain fraction of the optimal event rate for
NS-NS binary systems (cf. Eq.\,(\ref{Eq:SNR})). In a second step
these configurations are explored in the high frequency regime by
considering a smaller frequency integration interval $[500\,{\rm
Hz},1570\,{\rm Hz}]$ and selecting the optimal signal-to-noise
ratio on this interval. For the $(P^{(1)}= P^{(2)}=400\ {\rm
kW})$-configuration as an example, we can achieve a sensitivity
comparable to Advanced LIGO on the $[500\,{\rm Hz},1570\,{\rm
Hz}]$ frequency band while maintaining an improvement in the event
rate (integrating again from $f_{\rm min}$) of 16.4\% compared to
Advanced LIGO (cf. Fig\,\ref{Fig:NSD22}).

After adding up all classical noise contribution shown in
Figs.\,\ref{Fig:NSD12} and \ref{Fig:NSD22} it turns out that the
noise spectral density of the double optical spring configuration
-- contrary to the single optical spring in Advanced LIGO --
almost follows the borderline set by the classical noise in the
low frequency regime. Let us write
\begin{align}
\eta &\equiv \int_{f_{\rm min}}^{150\,{\rm Hz}}\frac{f^{-7/3}}{S_h (f)} \
{\rm d}f \Big/ \int_{f_{\rm min}}^{150\,{\rm
Hz}}\frac{f^{-7/3}}{S^{\rm cl}_h(f)} \ {\rm d}f \nonumber \\
&\equiv \frac{\bar S^{\rm cl}_h}{\bar S_h} = \frac{\bar S^{\rm cl}_h}{\bar S_h^{\rm cl}+\bar S_h^{\rm q}} \,,
\end{align}
where $\bar S^{\rm cl}_h$, $\bar S^{\rm q}_h$ and $\bar S_h$ are weighted averages of classical, quantum, and total noise spectrum in the low-frequency band, respectively. For the $P^{(1)}=P^{(2)}=400\ {\rm kW}$
configuration (cf. Fig.\,\ref{Fig:NSD22}) we obtain
\begin{equation}
\eta \approx
0.81\,.
\end{equation}
This indicates that, at low frequencies (i.e., below 150\,Hz),
the quantum noise is already small fraction of the total noise; improving quantum noise further does not significantly improve sensitivity. Qualitatively, starting with a level of $\eta = 0.81$, further lowering $\bar S^{\rm q}_h$ by a factor of 2 only improves $\eta$ from 0.81 to 0.88, which yields an 14\% increase in event rate.

\subsection{Optimized configurations with classical noise budget beyond Advanced LIGO}

It is likely that technical improvements will reduce the
classical noise floor in the future. In order to explore the
potential of our proposed configuration to increase the quantum
limited sensitivity by reshaping the noise curves in an optimal
way, we analyze the performance for a reduced classical noise
budget. For instance, the gravity gradient noise is a limiting
factor at lower frequencies. As suggested in Ref.~\cite{HuTh1998}
this effect can be removed from the recorded data by performing an
independent measurement of the ground's density fluctuations near
each test-mass. We assume it to be 1/10 (in amplitude) the current  estimation for Advanced LIGO~\cite{ADHIKARI}. Another limiting factor is
given by the thermal noise in the suspension system and mirrors.
We assume that the suspension thermal noise can be lowered by
a factor of 10 in amplitude, while internal thermal noise of mirrors
can be lowered by a factor of 3 in amplitude~\cite{ADHIKARI}.  Such improvements
may possibly be realized by (i) optimizing the design of mirror coating
structure and suspension wires, (ii) improving mechanical quality factors of mirror coating,
substrate and suspension materials, and (iii) applying cryogenic
techniques~\cite{GSSW1999,BBFDS2006,BoTh2006,UTTTMOKSSHYS1999,UTTTKSSYHS1998}.

Now we optimize the usual single optical spring configuration as
well as our proposed double optical spring layout for this
modified noise budget.
\begin{figure}[h!!!]
\includegraphics[height=\linewidth,angle=-90]{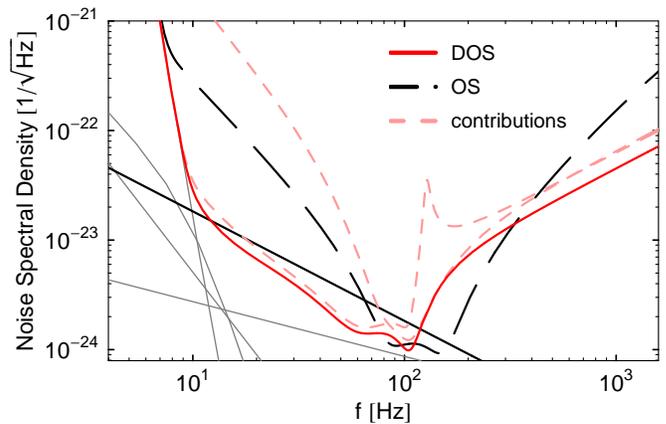}
\caption{Compared to Fig.\,\ref{Fig:NSD12} suspension thermal
noise and gravity gradient noise are lowered by a factor of 10 and
coating thermal noise by a factor of 3. DOS configuration as well
as single optical spring configuration both optimized with
respect to the new noise budget. An improvement in event rate of
238\% can be achieved by DOS configurations, compared with
61\% of single optical spring configurations.
The following
parameters were used: $P^{(1)}=P^{(2)}=400 \ {\rm kW}$,
$\lambda^{(1)}=2\pi \ 130 \ {\rm Hz}$, $\lambda^{(2)}=-2\pi \ 10 \
{\rm Hz}$, $\epsilon^{(1)}=2\pi \ 5 \ {\rm Hz}$,
$\epsilon^{(2)}=2\pi \ 45 \ {\rm Hz}$, $\zeta^{(1)}=2\pi \ 0.46$,
$\zeta^{(2)}=2\pi \ 0.3$. For single optical spring optimization, parameters are given in the first row of the second block in Tab.~\ref{Tab:improvement}.
}\label{Fig:lownoise}
\end{figure}
Only now the potential of the two carrier interferometer becomes
apparent: For the new classical noise budget, the single optical
spring configuration can improve the event rate for NS-NS binary
inspirals only by 61\%. But the DOS configuration can do 238\%.
This corresponds to an improvement in event rate of the DOS
compared to the optimized (with respect to the new classical noise
budget) single optical spring configuration of 110\% (cf.
Tab.\,\ref{Tab:improvement}).

Such a situation is presented in Fig.\,\ref{Fig:lownoise}. The big gap between the single optical spring
total noise and the classical noise budget verifies
that the single optical spring Advanced LIGO configuration has not
been limited by classical noise at low frequencies.
While this gap can be partially filled by the  double optical spring
configuration, there is still room for further
improvement.  For the configuration
used in Fig.\,\ref{Fig:lownoise} we evaluate
\begin{equation}
\eta \approx
0.3\,.
\end{equation}
In this case, lowering the quantum noise by a factor of 2 improves the event rate by 69\%. One possibility of further improvement would be to inject
even more than two carriers, and combine the corresponding
output channels (cf.~App.~\ref{sec:mos}).

\section{Conclusion}\label{sec:conclusion}
While the concept of a stable Double Optical Spring (DOS)
has motivated experiments such as the one already
carried out by Corbitt et al.~\cite{CCIMORSWWM2007}, in this
paper, we have theoretically investigated the benefit of a Double Optical
Spring (DOS) configuration for second generation gravitational-wave
detectors, in particular in a follow-up experiment to Advanced LIGO, possibly in
combination with other existing schemes, e.g., the local readout scheme proposed
in Ref.~\cite{RMSLSDC2007}, as well as the injection of
squeezed vacuum~\cite{McKenzie04,VCHFDS2006, HCCFVDS2003,BUCH2004,VCHFDS2005}.

In the DOS configuration, a second laser beam is injected
into signal-recycling interferometers at the bright port which is,
as the first carrier, resonant
in the arm-cavities and is also detuned in the signal-recycling
cavity. The two outputs are optimally filtered and combined. By
choosing appropriate detunings of the signal-recycling cavity and
homodyne detection angles, it is possible to achieve a stable
double optical spring while additionally improving the
sensitivity.

Taking into account the current classical noise budget estimation
for the Advanced LIGO detector, as well as constraints on optical
power, we have performed an optimization of our double-carrier
scheme towards the detection of compact binary inspirals
specifying to neutron stars. The DOS allows a 36\% improvement
in event rate, and we have shown that further improvement
in event rate will largely be limited by classical noise. When
considering a more optimistic classical noise budget, DOS
interferometers are much more capable than single-optical-spring
interferometers in taking advantage of this improvement: compared with
61\% improvement in event rate achievable by single-optical-spring configurations,
the DOS allows 238\%.  Nevertheless the reduced classical noise level leaves further room beyond
DOS, which can be exploited by employing more than two carrier fields.

\begin{acknowledgments}
We thank S.~Waldman and K.~Strain as well as T.~Corbitt, N.~Mavalvala and C.~Wipf for very useful discussions.
We thank R.~Adhikari for suggesting the level of reduction of
classical noise. Research of H.M.-E.\,, K.S.\, and Y.C.\, is
supported by the Alexander von Humboldt Foundation's Sofja
Kovalevskaja Programme. Y.C. is also supported by NSF grants
PHY-0653653 and PHY-0601459, as well as  the David and Barbara Groce
startup fund at Caltech.  Research of S.L.D. is also supported by
the Alexander von Humboldt Foundation. Research of H.R.\, and
R.S.\, is supported by the Deutsche Forschungsgemeinschaft through
the SFB No. 407.
\end{acknowledgments}

\begin{appendix}
\section{Optimal output channel for interferometers with multiple carriers}
\label{sec:mos}
Here we analyze the optimal way of combining the output channels obtained
from homodyne detections at the dark port of a multi-carrier interferometer. The general equation of motion for the mirror position in case of $n$ carriers reads [cf.~Eq.~\eqref{eomx}]:
\begin{equation}
\hat x = - R_{xx}(\Omega)\sum_{i=1}^{n}\left[\hat
F^{(i)}+R_{FF}^{(i)}(\Omega) \ \hat x\right] + L\ h + \hat
\xi_{\rm noise}\,.
\end{equation}
The output corresponding to each of the $n$ carrier fields is
given by [cf.~Eqs.~\eqref{eom1} and \eqref{eom2}]:
\begin{align}
\hat y^{(i)}(\Omega) = & \hat Y_1^{(i)} (\Omega) \ \sin \zeta^{(i)} + \hat Y_2^{(i)} (\Omega) \ \cos \zeta^{(i)}\notag\\
& + [R_{Y_1F}^{(i)} (\Omega) \ \sin \zeta^{(i)} + R_{Y_2F}^{(i)}
(\Omega) \ \cos \zeta^{(i)}] \hat x\notag\\
 \equiv & \vec{n}_i^T\vec{\nu}+s_i h
\end{align}
where $\vec{\nu}$ is a vector with $2n+1$ entries which account for $2n$ quadrature operators and one operator modeling the classical noise. Here $\vec{n}_i$ describes the noise transfer functions from the noise channels $\vec{\nu}$ into the output channels and $s_i$ accounts for the signal-transfer functions. The combined output is given by
\begin{equation}
\hat y(\Omega) = \sum_{i=1}^n K_i (\Omega) \ \hat y^{(i)}(\Omega)\,,
\end{equation}
and one has to identify $n$ optimal filter functions which
minimize the signal-referred noise spectral density of $\hat y $:
\begin{equation}
S_h(\Omega)=\frac{\sum_{i,j=1}^n (\mathbf{N})_{ij}K_i K^*_j}{\sum_{i,j=1}^n (\mathbf{S})_{ij} K_i K^*_j}
\end{equation}
with
\begin{equation}
(\mathbf{S})_{ij}=s_i s_j^*
\end{equation}
and
\begin{equation}
(\mathbf{N})_{ij} = \sum_{s,k=1}^{2n+1}(\vec{n}_i^T)_s
S_{\nu_s \nu_k} (\vec{n}_j^\dag)_k \,.
\end{equation}
Here $S_{\nu_s \nu_k}$ is the cross spectral density between
$\nu_s$ and $\nu_k$, with
\begin{equation}
\langle \nu_s (\Omega)\nu_k^\dagger(\Omega') \rangle  = 2\pi\delta(\Omega-\Omega') \frac{S_{\nu_s \nu_k}(\Omega)}{2}
\end{equation}

The inverse of the largest eigenvalue of the $n\times n$-matrix
\begin{equation}
\mathbf{M}=\mathbf{N}^{-1}\cdot\mathbf{S}\,.
\end{equation}
provides the resulting minimum noise and the corresponding eigenvector gives the $n$ optimal filter function $K_i$.
\end{appendix}

\end{document}